\documentclass[twocolumn,showpacs,preprintnumbers,amsmath,amssymb]{revtex4}
\usepackage{graphicx}% Include figure files
\usepackage{dcolumn}% Align table columns on decimal point
\usepackage{bm}% bold math
\usepackage{subfigure}
\usepackage[colorlinks,linkcolor=blue,anchorcolor=blue,citecolor=blue]{hyperref}

\begin{document}

%\preprint{APS/123-QED}

\title{Quantum Dimensionality Reduction by Linear Discriminant Analysis}% Force line breaks with \\

\author{Kai  Yu$^{1}$}
\author{Gong-De Guo$^{1}$}%
\author{Song Lin$^{1}$}%
\altaffiliation{Corresponding author. Email address: lins95@gmail.com}
 %\altaffiliation[Also at ]{ College of Mathematics and Informatics, Fujian Normal University, Fuzhou 350117, China}%Lines break automatically or can be forced with \\

\affiliation{%
College of Mathematics and Informatics, Fujian Normal University, Fuzhou 350117, China
}%

%\author{song Li}
%\affiliation{

\date{\today}% It is always \today, today,

\begin{abstract}
Dimensionality reduction (DR) of data is a crucial issue for many machine learning tasks, such as pattern recognition and data classification. In this paper, we present a quantum algorithm and a quantum circuit to efficiently perform linear discriminant analysis (LDA) for dimensionality reduction. Firstly, the presented algorithm improves the existing quantum LDA algorithm to avoid the error caused by the irreversibility of the between-class scatter matrix $S_B$ in the original algorithm. Secondly, a quantum algorithm and quantum circuits are proposed to obtain the target state corresponding to the low-dimensional data. Compared with the best-known classical algorithm, the quantum linear discriminant analysis dimensionality reduction (QLDADR) algorithm has exponential acceleration on the number $M$ of vectors and a quadratic speedup on the dimensionality $D$ of the original data space, when the original dataset is projected onto a polylogarithmic low-dimensional space. Moreover, the target state obtained by our algorithm can be used as a submodule of other quantum machine learning tasks. It has practical application value of make that free from the disaster of dimensionality.
\end{abstract}

\maketitle

\section{\label{sec1}Introduction}

\par Nowadays, we are living in an era of big data. Hence, the processing capacity of big data is increasingly required in this world. As we all know, with the increase of data dimension, high-dimensional data usually has greater coherence and redundancy. Moreover, the information content of data itself grows more slowly than the data dimension. That is to say, the higher the signal dimension is, the greater the data redundancy will be. To overcome the influence of high-dimensional data \cite{ref:CMB,ref:GA}, a method is proposed to map high-dimensional data to low-dimensional data by utilizing the sparsity and redundancy of high-dimensional data, which named dimensionality reduction. At present, a series of techniques have been put forward for dimensionality reduction. For instance, principal component analysis (PCA) dimensionality reduction technique is guaranteed in terms of the maximum retention data variances \cite{ref:JL}.
\par Different from the aforementioned PCA which is not considered the data category, Fisher proposed a novel dimensionality reduction algorithm named linear discriminant analysis (LDA) \cite{ref:CMB}.  This approach can project the data in the direction where maximizes the between-class variance but minimizes the within-class variance. It is not surprising that LDA is shown to more effective than PCA in machine learning involving classification \cite{ref:PNB,ref:YQC}.
\par However, the classical LDA algorithm faces the same problem as other classical machine learning algorithms, namely, high time complexity. To optimize it, a range of quantum algorithms have been proposed in machine learning, which achieved exponential acceleration compared with the classical ones \cite{ref:JA,ref:PW}. In particular, quantum algorithms for solving the problem of pattern classification and regression analysis problems were proposed, covering an important area of machine learning \cite{ref:MS,ref:BJD,ref:CHY}. Recently, the quantum neural network algorithms which combines quantum information theory and artificial neural networks \cite{ref:JZ}, variational quantum algorithms \cite{ref:AP,ref:MC,ref:RL,ref:YL} are proposed, let us see the potential application prospect of quantum machine learning algorithms in the age of big data.
\par In the application field of quantum dimensionality reduction, the quantum algorithm for PCA has been proposed for unsupervised mode \cite{ref:SL,ref:CHY2}, the quantum algorithm for A-optimal projection is used in regression tasks \cite{ref:BJD2}, and Cong et al. gave the quantum LDA algorithm \cite{ref:IC}. However, Cong et al. only prepared a set of principal component vectors that can indirectly obtain the optimal projection direction in parallel, not obtain the quantum states corresponding to the principal components and the low-dimensional data vectors.
\par In this paper, we present a quantum linear discriminant analysis dimensionality reduction (QLDADR) algorithm. The algorithm focuses on the vectors in which a high-dimensional feature space are projected onto a low-dimensional feature space. Moreover, it could generate a target state corresponding to the reduced dimensional data, to service other quantum algorithms. The analysis shows the proposed algorithm has exponential acceleration in the number of input data and quadratic acceleration in the data dimension compared with the classical algorithm.
\par The paper is organized as follows. We give a brief overview of classical LDA algorithm in Sec. \ref{sec:2}. In Sec. \ref{sec:3}, the quantum LDA dimensionality reduction algorithm and quantum circuits are presented. A brief analysis of this quantum algorithm is showed in Sec. \ref{sec:4}. Finally, a short conclusion is provided in Sec. \ref{sec:5}.

\section{\label{sec:2}Review of Classical LDA Dimensionality Reduction Algorithm}
\par LDA is a popular dimensionality reduction algorithm in machine learning. Now, we review the basic idea of  the classical LDA dimensionality reduction algorithm as follows.
\par Considering a dataset $\{\mathbf{x}_{i}\in \mathbb{R}^{D}:1\leq i\leq M\}$, and each $\mathbf{x}_{i}$ is represented by a $D$-dimensional column vector $\mathbf{x}_{i}={({x}_{i1},{x}_{i2},\cdots,{x}_{iD})}^{T}$. Furthermore, we assume that each data vector $\mathbf{x}_{i}$ in dataset $\{\mathbf{x}_{i}\}_{i=1}^{M}$ has been divided into one of $n$ categories. For such a data set, LDA projects it into a low-dimensional space to maximize the between-class variance (for class differentiation) while minimizing the within-class variance.
\par Let $\boldsymbol{\mu}_{c}$ denotes the within-class mean (centroid) of class $c(1\leq c\leq n)$, and $\bar{\boldsymbol{o}}$ is represented the mean of all data points $\mathbf{x}$. Then the within-class scatter matrix can be expressed as
     \begin{equation}
        S_{W}=\sum_{i=1}^{M}(\mathbf{x}_{i}-\boldsymbol{\mu}_{c_i})(\mathbf{x}_{i}-\boldsymbol{\mu}_{c_i})^{T}.
     \label{eq:1}
     \end{equation}
Here, $c_{i}$ is just a label for the class to which the data vector $\mathbf{x}_{i}$ belongs, eg., $c_{p}=c_{q}=c$ if both $\mathbf{x}_{p}$ and $\mathbf{x}_{q}$ are in category $c$. Furthermore, the between-class scatter matrix can be expressed as
     \begin{equation}
        S_{B}=\sum_{c=1}^{n}(\boldsymbol{\mu}_{c}-\bar{\boldsymbol{o}})(\boldsymbol{\mu}_{c}-\bar{\boldsymbol{o}})^{T}.
     \label{eq:2}
     \end{equation}
\par The goal of the LDA algorithm is to find a direction $\boldsymbol{\omega} \in \mathbb{R}^{D}$, and the projection of any data vector in this direction is $\boldsymbol{\omega}^{T}\mathbf{x}_{i}$. Of course, the LDA algorithm must maximize the between-class variance $\boldsymbol{\omega}^{T}S_{B}\boldsymbol{\omega}$ and minimize the within-class variance $\boldsymbol{\omega}^{T}S_{W}\boldsymbol{\omega}$. Therefore, the objective function of the algorithm can be expressed as
     \begin{equation}
        \max _{\boldsymbol{\omega}} J(\boldsymbol{\omega})=\frac{\boldsymbol{\omega}^{T} S_{B} \boldsymbol{\omega}}{\boldsymbol{\omega}^{T} S_{W} \boldsymbol{\omega}}.
     \label{eq:3}
     \end{equation}
Noticed that the numerator and denominator of Eq. (\ref{eq:3}) are quadratic terms related to $\boldsymbol{\omega}$. Hence, the solution to $J(\boldsymbol{\omega})$ only has influence of the direction of $\boldsymbol{\omega}$, rather than, the length of $\boldsymbol{\omega}$. Without loss of generality, assuming that $\boldsymbol{\omega}^{T}S_{W}\boldsymbol{\omega}=1$ is satisfied, then the optimization objective is equivalent to
     \begin{equation}
        \begin{split}
            \min _{\boldsymbol{\omega}} & -{\boldsymbol{\omega}^{T} S_{B} \boldsymbol{\omega}}\\
            {\rm subject} \ {\rm to} & \ {\boldsymbol{\omega}^{T} S_{W} \boldsymbol{\omega}=1}.
        \end{split}
     \label{eq:4}
     \end{equation}
According to the method of Lagrangian Multiplier, the Lagrangian function corresponding to the optimization objective is
     \begin{equation}
        F(\boldsymbol{\omega},\lambda)=-{\boldsymbol{\omega}^{T} S_{B} \boldsymbol{\omega}}+\lambda({\boldsymbol{\omega}^{T} S_{W} \boldsymbol{\omega}-1}),
     \label{eq:5}
     \end{equation}
where $\lambda$ is the desired Lagrange multiplier. By solving the partial derivative of the variable, we get
     \begin{equation}
        {S_{W}^{-1} S_{B} {\boldsymbol{\omega}}}=\lambda{\boldsymbol{\omega}},
     \label{eq:6}
     \end{equation}
which shows that $\boldsymbol{\omega}$ is an eigenvector of $S_{W}^{-1} S_{B}$.
\par In order to project the original dataset to a lower-dimensional space of dimension $d$, it needs to seek $d$ vectors $(\boldsymbol{\omega}_{1},\boldsymbol{\omega}_{2},\cdots,\boldsymbol{\omega}_{d})$, which form a basis for the projection subspace. These vectors make up a $D\times d$ matrix $W$. In this case, corresponded to maximizing the discriminant
     \begin{equation}
        \max _{W} J(W)=\frac{W^{T} S_{B} W}{W^{T} S_{W} W}.
     \label{eq:7}
     \end{equation}
\par Moreover, we can easily prove the column of $W$ will be the eigenvector corresponding to the $d$ maximum eigenvalues of $S_{W}^{-1} S_{B}$, just as in the case of principal component analysis. Hence, the data set after dimensionality reduction can be expressed as
     \begin{equation}
        Y=X_{M \times D}{W_{D \times d}}.
     \label{eq:8}
     \end{equation}
It's easy to find that $Y$ is a matrix of $M \times d$, and the $i$th row can be written as
     \begin{equation}
        \mathbf{y}_{i}^{T}=\mathbf{x}_{i}^{T}{W},
     \label{eq:9}
     \end{equation}
where $\mathbf{y}_{i}=({y}_{i1},{y}_{i2},\cdots,{y}_{id})^{T} \in \mathbb{R}^{d}$.

\section{\label{sec:3}Quantum Data Dimensionality Reduction by LDA Algorithm}
\par Quantum computer provides a new platform to solve the problem of dimensionality reduction. Quantum random access memory (QRAM) gives an architecture that exponentially reduces the requirement for memory cells to store vectors \cite{ref:VG}. We assume that each data vector is stored in QRAM in terms of its difference from the class means. That is, if a data vector $\mathbf{x}_{i}$ belongs to class $c(1\leq c \leq n)$ with centroid $\boldsymbol{\mu}_{i}$, then the data vector $\mathbf{x}_{c_{i}}$, the class label $c_i$ (index $i$ is just a mark for the class of the $i$th data $\mathbf{x}_{i}$) and the centroid $\boldsymbol{\mu}_{i}$ are stored as floating-point numbers in QRAM \cite{ref:SL2}. Furthermore, the above message are stored in the data structure proposed in Ref. \cite{ref:IK}, which allows us to efficiently perform the following two unitary operations.
%eq:10 UD
     \begin{equation}
        U_{\mathcal{D}}:|i\rangle|0\rangle \rightarrow \frac{\sum_{j=1}^{D} x_{i j}|i\rangle|j\rangle}{\left\|\mathbf{x}_{i}\right\|},
     \label{eq:10}
     \end{equation}
%eq:11 UM
     \begin{equation}
        U_{\mathcal{M}}:|0\rangle|j\rangle \rightarrow \frac{\sum_{i=1}^{M} {\left\|\mathbf{x}_{i}\right\|}|i\rangle|j\rangle}{{\left\|X\right\|}_{F}},
     \label{eq:11}
     \end{equation}
where ${\left\|\mathbf{x}_{i}\right\|}$ denotes $2$-norms of $\mathbf{x}_{i}$ and ${{\left\|X\right\|}_{F}}$ is Frobenius norm of $X$. Then, we use the two unitary operations $U_{\mathcal{D}}$ and $U_{\mathcal{M}}$ to generate the desired initial state corresponding to the original dataset,
     \begin{equation}
        \begin{split}
            |\psi_{X}\rangle &= U_{\mathcal{D}}U_{\mathcal{M}}|0\rangle|0\rangle\\
            &=U_{\mathcal{D}}\frac{\sum_{i=1}^{M} {\left\|\mathbf{x}_{i}\right\|}|i\rangle|0\rangle}{{\left\|X\right\|}_{F}}\\
            &=\frac{\sum_{i=1}^{M}\sum_{j=1}^{D} x_{i j}|i\rangle|j\rangle}{{\left\|X\right\|}_{F}},
        \end{split}
     \label{eq:12}
     \end{equation}
in time $O\left( ploylog(MD) \right)$.
%Basic Ideas of QLDADR
\subsection{\label{sec:3.1}Basic Ideas of QLDADR}
\par In the context of quantum information, the task of dimensionality reduction means the quantum state $|\psi_{X}\rangle$ mapped to quantum state $|\psi_{Y}\rangle$ that denotes a state of the low-dimensional dataset. It can be described as follows,
     \begin{equation}
        \begin{split}
            |\psi_{X}\rangle &=\frac{\sum_{i=1}^{M}\sum_{j=1}^{D} x_{i j}|i\rangle|j\rangle}{{\left\|X\right\|}_{F}}\\
           \rightarrow \left|\psi_{Y}\right\rangle &=\frac{\sum_{i=1}^{M} \sum_{j=1}^{d} y_{i j}|i\rangle|j\rangle}{\|Y\|_{F}},
        \end{split}
     \label{eq:13}
     \end{equation}
where ${\|Y\|_{F}}$ denotes Frobenius norm of  $Y$. With the help of the following steps, the task can then be solved more efficiently.
%Extract the shadow principal components
\subsubsection{\label{sec:3.1.1}Extract the shadow principal components}
\par According to the Sec. \ref{sec:2}, we need to find the principal components to help accomplish the task of dimensionality reduction by LDA. By Eq. (\ref{eq:6}), our main task is to solve the eigenvector problem of $S_{W}^{-1}{S_{B}}$. Obviously, this problem would be simple if only $S_{W}^{-1}{S_{B}}$ is Hermitian positive semidefinite.
\par To simplify this problem, we turn it into a density matrix problem. Specifically, preparing a Hermitian positive definite matrix $S_{W}$, and letting $\boldsymbol{\omega}={S_{W}^{-{1}/{2}}}{\mathbf{v}}$. Then, the problem is reduced to the following eigenvalue problem
     \begin{equation}
        {S_{W}^{-{1}/{2}}}{S_{B}}{S_{W}^{-{1}/{2}}}{\mathbf{v}}={\lambda}{\mathbf{v}},
     \label{eq:14}
     \end{equation}
where $\lambda$ is a eigenvalue of ${S_{W}^{-{1}/{2}}}{S_{B}}{S_{W}^{-{1}/{2}}}$ and corresponding a eigenvector ${\mathbf{v}}$. In addition, it is easy to find that ${S_{W}^{-{1}}}{S_{B}}$ and ${S_{W}^{-{1}/{2}}}{S_{B}}{S_{W}^{-{1}/{2}}}$ have the same eigenvalues by observing Eq. (\ref{eq:6}) and Eq. (\ref{eq:14}). And ${\boldsymbol{\omega}}$ can be obtained from ${\mathbf{v}}$, so we regard ${\mathbf{v}}$ as the shadow of ${\boldsymbol{\omega}}$.
\par Through the above analysis, it isn't hard to notice that way with the help of ${\mathbf{v}}$ is a good strategy to achieve dimensionality reduction task based on LDA. Therefore, the first task is transformed to extract the eigenvectors' quantum form $|{\mathbf{v}_{1}}\rangle,|{\mathbf{v}_{2}}\rangle,\cdots,|{\mathbf{v}_{d}}\rangle$ corresponding to the first $d$ maximum eigenvalues, which are named shadow principal components. The details of preparing $\{|{\mathbf{v}_{j}}\rangle\}_{j=1}^{d}$ are shown as follows.
%\begin{enumerate}
%   \item[(S1.1)] Firstly, initializing the Hermitian operators $S_W$ and $S_B$ in the way of Appendix \ref{appendix A}. We %then utilize the technique of Implementing the Hermitian chain product in \cite{ref:IC} to construct a density matrix
%\end{enumerate}
\par (S1.1) Firstly, initializing the Hermitian operators $S_W$ and $S_B$ in the way of Appendix \ref{appendix A}. We then utilize the technique of \emph{Implementing the Hermitian chain product } in \cite{ref:IC} to construct a density matrix
     \begin{equation}
        \rho={S_{W}^{-{1}/{2}}}{S_{B}}{S_{W}^{-{1}/{2}}}=\sum_{i=1}^{M}{\lambda_{j}}{|{\mathbf{v}_{j}}\rangle\langle{\mathbf{v}_{j}}|},
     \label{eq:15}
     \end{equation}
in time $O\left(log(MD){k_{\lambda}^{3.5}} / {{\epsilon}_{\lambda}^{3}}\right)$. The $\lambda_{j}$ and $\mathbf{v}_{j}$ are the eigenvalues and eigenvectors of $S_W^{-1/2} S_B S_W^{-1/2}$, respectively. To avoid exponential complexity in the case of exponentially small eigenvalues, we use a method of \cite{ref:PR} to pre-define an effective condition number $k_{\lambda}$ and making the eigenvalues of the phase estimation within the range $[1/ k_{\lambda},1]$. Without loss of generality, it is assumed that the eigenvalues have been arranged in descending order, that is, $\lambda_{1}\geq \lambda_{2} \geq \cdots\geq\lambda_{D}\geq 0$.
\par (S1.2) Then, we construct an unitary operator $e^{-{\rm i}\rho t}$ $({\rm i}^{2}={-1})$ by using the matrix exponentiation technique presented in \cite{ref:SL}. Then, implementing quantum phase estimation \cite{ref:MAN} on $\rho$, which the controlled unitary gate is $e^{-{\rm i}\rho t}$. In this way, we can obtain an approximation to the state
     \begin{equation}
        \rho^{\prime}=\sum_{j=1}^{D} \lambda_{j}\left|\mathbf{v}_{j}\right\rangle\left\langle \mathbf{v}_{j}|\otimes| \lambda_{j}\right\rangle\left\langle\lambda_{j}\right|.
     \label{eq:16}
     \end{equation}
\par (S1.3) Finally, measuring the system of $| \lambda_{j} \rangle$ for $O(d)$ times. Then, we get the first $d$ maximum eigenvalues $\lambda_{1},\lambda_{2},\cdots,\lambda_{d}$ with high probability. At same time, we can obtain the quantum states $|\mathbf{v}_{j} \rangle$ of eigenvectors, corresponding to the eigenvalues $\lambda_{j}$. According to \cite{ref:JY}, the maximum value of $d$ is generally set as $n-1$, where $n$ is the number of categories of the original data set. Furthermore, the value of $d$ satisfies
     \begin{equation}
        \min _{d} \sum_{j=1}^{d} \lambda_{j} \geq \mathcal{S},
     \label{eq:17}
     \end{equation}
by \cite{ref:CHY2}. That is to say, the cumulative variance sum of the first $d$ principal components is greater than a preset threshold value $\mathcal{S}$, and the threshold value is close to $1$.
%Prepare the intermediate state
\subsubsection{\label{sec:3.1.2}Prepare the intermediate state}
\par Consider the shadow principal component $\mathbf{v}$ is introduced for auxiliary projection dimension reduction. Therefore, we need to prepare a quantum state $|\psi_{T} \rangle$ that can effectively interact with the quantum state $|\mathbf{v}_{j} \rangle$ on the basis of $|\psi_{X} \rangle$, which is called the intermediate quantum state. The detailed steps are as follows.
\par (S2.1) Our algorithm aims to obtain the target state $|\psi_{Y} \rangle$. Now, we rewrite the original data vector $\mathbf{x}_{i}$ on the basis $\{|{\boldsymbol{\omega}_{1}}\rangle,|{\boldsymbol{\omega}_{2}}\rangle,\cdots,|{\boldsymbol{\omega}_{D}}\rangle\}$ as
     \begin{equation}
        \mathbf{x}_{i}=\left( {\sum_{j=1}^{D}|\boldsymbol{\omega}_{j}\rangle\langle \boldsymbol{\omega}_{j}|}\right){\mathbf{x}_{i}}=\sum_{j=1}^{D}{y_{ij}}{\left|\boldsymbol{\omega}_{j}\right\rangle}.
     \label{eq:18}
     \end{equation}
Moreover, using the two unitary operations $U_{\mathcal{D}}$ and $U_{\mathcal{M}}$ to construct the $|\psi_{X} \rangle$ . It can be mathematically reformulate as
     \begin{equation}
        \begin{split}
            |\psi_{X}\rangle &= \frac{\sum_{i=1}^{M}\sum_{j=1}^{D} x_{i j}|i\rangle|j\rangle}{{\left\|X\right\|}_{F}}\\
            &=\frac{\sum_{i=1}^{M}|i\rangle \mathbf{x}_{i}}{{\left\|X\right\|}_{F}}\\
            &=\frac{\sum_{i=1}^{M}\sum_{j=1}^{D} y_{i j}|i\rangle|\boldsymbol{\omega}_{j}\rangle}{{\left\|X\right\|}_{F}}.
        \end{split}
     \label{eq:19}
     \end{equation}
\par (S2.2) Assuming there is a function $f(X)=X^{1/2}$ with convergent Taylor series \cite{ref:IC}. Thus, we can obtain a Hermitian operator $S_{W}^{1 / 2}=\sum_{k=1}^{D} \sigma_{k}\left|\mathbf{u}_{k}\right\rangle\left\langle \mathbf{u}_{k}\right|$ by executing $f(S_{W})$. Taking $O(\epsilon_{\sigma}^{-3})$ copies of state $S_{W}^{1/2}$ to produce an unitary operation $e^{-{\rm i}S_{W}^{1/2}t}$ by using the technique of \cite{ref:SL}. Then, we apply quantum phase estimation to estimate its eigenvalues with the error $\epsilon_{\sigma}$. In this case, we append a register of $log(\sigma_{k}^{-1})$ qubits in the state $|0\rangle$ to estimate the eigenvalues of $e^{-{\rm i}S_{W}^{1/2}t}$, and combine the second register $|\boldsymbol{\omega}_{j}\rangle$ of state $|\psi_{X}\rangle$ in Eq. (\ref{eq:19}). After the above operations, the approximate state of the whole system is
     \begin{equation}
        \left|\Phi_{1}\right\rangle=\frac{\sum_{i=1}^{M} \sum_{j,k=1}^{D} y_{i j}\beta_{j k}|i\rangle\left|\mathbf{u}_{k}\right\rangle\left|\sigma_{k}\right\rangle}{\|X\|_{F}},
     \label{eq:20}
     \end{equation}
where $\beta_{j k}=\left\langle \mathbf{u}_{k} | \boldsymbol{\omega}_{j}\right\rangle$.
\par (S2.3)  Next, appending a qubit in the state $|0\rangle$ as the last register and rotating it to $\sqrt{1-C_{1}^{2} \sigma_{k}^{2}}|0\rangle+C_{1} \sigma_{k}|1\rangle$ by $|\sigma_{k}\rangle$ controlled, where $C_{1}=O(max_{k}\sigma_{k}^{-1})$. The results in the overall state
     \begin{equation}
        \begin{split}
           & \left|\Phi_{2}\right\rangle=\\
           &\frac{{\sum_{i=1}^{M} \sum_{j,k=1}^{D} y_{i j}\beta_{j k}|i\rangle\left|\mathbf{u}_{k}\right\rangle\left|\sigma_{k}\right\rangle}{\left(\sqrt{1-C_{1}^{2} \sigma_{k}^{2}}|0\rangle+C_{1} \sigma_{k}|1\rangle\right)}}{\|X\|_{F}}.
        \end{split}
     \label{eq:21}
     \end{equation}
\par (S2.4) Applying inverse phase estimation to undo step (S2.2) and discarding the register $|\sigma_{k}\rangle$. The remaining registers are in state
     \begin{equation}
        \begin{split}
           & \left|\Phi_{3}\right\rangle=\\
           &\frac{{\sum_{i=1}^{M} \sum_{j,k=1}^{D} y_{i j}\beta_{j k}|i\rangle\left|\mathbf{u}_{k}\right\rangle}{\left(\sqrt{1-C_{1}^{2} \sigma_{k}^{2}}|0\rangle+C_{1} \sigma_{k}|1\rangle\right)}}{\|X\|_{F}}.
        \end{split}
     \label{eq:22}
     \end{equation}
\par (S2.5)  A projective measurement $|1\rangle \langle 1|$ is performed on the last register of state of Eq. (\ref{eq:22}), then the outcome is in the state $|1\rangle$ with probability $p_{1}=\frac{{\sum_{i=1}^{M} \sum_{j,k=1}^{D} y_{i j}^{2}\beta_{j k}^{2}}{C_{1}^{2}\sigma_{k}^{2}}}{\|X\|_{F}^{2}}$. If the measurement succeed, we have state of other registers,
     \begin{equation}
        \begin{split}
          |\psi_{T}\rangle &=\frac{{\sum_{i=1}^{M} \sum_{j,k=1}^{D} y_{i j}\beta_{j k}{C_{1}}{\sigma_{k}}|i\rangle\left|\mathbf{u}_{k}\right\rangle}}{\|X\|_{F} \sqrt{p_{1}}}\\
           &=\frac{\sum_{i=1}^{M}\sum_{j=1}^{D}{y_{i j}}{C_{1}}{|i\rangle}{\left|\mathbf{v}_{j}\right\rangle}}{\|X\|_{F} \sqrt{p_{1}}}.
        \end{split}
     \label{eq:23}
     \end{equation}
which called the intermediate state.
\subsubsection{\label{sec:3.1.3}Prepare the unitary operation $U(\lambda_{j})$}
\par
\begin{figure}
        %\centering
        \includegraphics[width=0.46\textwidth]{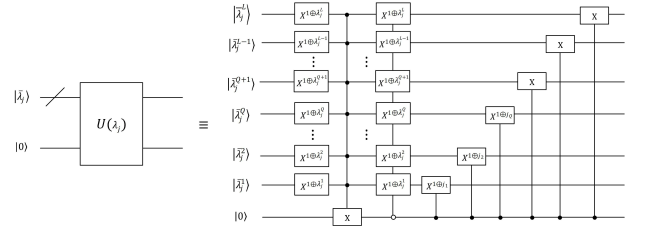}% Here is how to import EPS art
        \caption{Quantum circuit for $U(\lambda_{j})$, where the upper register stores the eigenvalues $|\bar{\lambda}_{j}\rangle$ for $j=1,2,\cdots,d$, and the lower register is the signal register. The state $|\bar{\lambda}_{j}\rangle$ of eigenvalues can be represented by $|\lambda_{j}^{L}\lambda_{j}^{L-1}\cdots \lambda_{j}^{2}\lambda_{j}^{1}\rangle $. Here, $X$ denotes quantum NOT gate and the label $Q \leq L$. Furthermore, $X^{1 \oplus \lambda_{j}^{l}}=I(X)$ if $\lambda_{j}^{l}=1 (0)$. It means that $\bar{\lambda}_{j}=\lambda_{j}$ if $\bar{\lambda}_{j}^{l}=\lambda_{j}^{l}$ for all $l=1,2,\cdots,L$. In this case, the last register implements the map $|0\rangle\mapsto |1\rangle$, and then implements $CU(X^{1\oplus{j_{q}}})$ to achieve the map $|{\bar{\lambda}_{j}}\rangle\mapsto |j\rangle$. (Here, executing the unitary operation $X^{1\oplus{j_{q}}}$ when the last register is in the state $|1\rangle$; and $|j\rangle=|0^{\otimes{L-Q}} j_{Q} j_{Q-1} \cdots j_{2} j_{1}\rangle$). On the contrary, if $\bar{\lambda}_{j}^{l} \neq \lambda_{j}^{l}$ exists, it means that $|\bar{\lambda}_{j}\rangle$ is not in the quantum state corresponding to the first $d$ maximum eigenvalues, and the circuit is equivalent to not doing any operation.}
    \label{fig:1}
    \end{figure}

 Distinguish between principal and non-principal components, we need to prepare a unitary operation $U(\lambda_{j})$. The idea of branching is whether the eigenvalue $\bar{\lambda}_{j}$ belongs to the first $d$ maximum eigenvalues $\{\lambda_{j}\}_{j=1}^{d}$ which obtained in step (S1.3). Moreover, each eigenvalue has the binary representation $\lambda_{j}=\lambda_{j}^{L}\lambda_{j}^{L-1}\cdots \lambda_{j}^{2}\lambda_{j}^{1}$. In this case, we construct a new unitary gate $X^{1\oplus \lambda_{j}^{l}}$ based on $X$ gate, to achieve the following function.
     \begin{equation}
        X^{1 \oplus \lambda_{j}^{l}}\left|\bar{\lambda}_{j}^{l}\right\rangle=
        \begin{cases}
            1, & \lambda_{j}^{l}=\bar{\lambda}_{j}^{l} \\
            0, & \lambda_{j}^{l} \neq \bar{\lambda}_{j}^{l}
        \end{cases},
     \label{eq:24}
     \end{equation}
where $j=1,2,\cdots,d$ is the index of eigenvalues, $l=1,2,\cdots,L$ represents the $l$th binary bit of a eigenvalue and $\lambda_{j}^{l}(\bar{\lambda}_{j}^{l})\in \{0,1\}$. Furthermore, $X^{1 \oplus \lambda_{j}^{l}}$ can be applied as the primitive to construct more complex unitary operations $U(\lambda_{j})$ acting as
     \begin{equation}
        U(\lambda_{j}):
        \begin{cases}
            |\bar{\lambda}_{j}\rangle |0\rangle\mapsto|j\rangle|1\rangle, &\bar{\lambda}_{j}=\lambda_{j}\\
            |\bar{\lambda}_{j}\rangle |0\rangle\mapsto|\bar{\lambda}_{j}\rangle |0\rangle, &\bar{\lambda}_{j} \neq \lambda_{j}
        \end{cases},
     \label{eq:25}
     \end{equation}
where the second register $|0\rangle$ is the signal register. And the state $|0\rangle$ is flipped to $|1\rangle$ only if $\bar{\lambda}_{j}=\lambda_{j}$. The quantum circuit for $U(\lambda_{j})$ is shown in Fig. \ref{fig:1}.
%Quantum LDA Dimensionality Reduction
\subsection{\label{sec:3.2}Quantum LDA Dimensionality Reduction}
\par In this section, we further design our algorithm based on the above ideas to obtain the target state $|\psi_{Y} \rangle$, which corresponded to the data set after dimensionality reduction. The detailed process of the construction consists of the following steps:
\par (1) \emph{Extract the shadow principal components}. Now, we extract the $d$ shadow principal components with the idea of Sec. \ref{sec:3.1.1}. First, constructing the density operator $\rho$ by querying QRAM and using the technique of \emph{Implementing the Hermitian chain product } in \cite{ref:IC}.  Then, applying the matrix exponentiation technique on $O(1/\epsilon_{\lambda}^{3})$ copies of $\rho$ to generate $e^{-{\rm i}\rho t}$ $({\rm i}^{2}={-1})$. Finally, using quantum phase estimation and sampling from the resulting probabilistic mixture, we can obtain the first d eigenvalues $\lambda_{j}$ in time $O\left( d ploylog(M D) k_{\lambda}^{3.5}/\epsilon_{\lambda}^{3} \right)$. Apparently, the shadow principal component $|\mathbf{v}_{j}\rangle$ are easy to be extracted, which correspond to the $j$th largest eigenvalue for$j=1,2,\cdots,d$.
\par (2) \emph{Prepare the intermediate state}. Following the ideas of Sec. \ref{sec:3.1.2}, we prepare the intermediate state $| \psi_{T}\rangle$ based on $| \psi_{X}\rangle$, which can effectively interact with $|\mathbf{v}_{j}\rangle$. The general flow of this is as follows (see Sec. \ref{sec:3.1.2} for more details):
\par (2.1) Using unitary operations $U_\mathcal{D}$ and $U_\mathcal{M}$ to query the QRAM, constructing the initial state $|\psi_{X}\rangle$ in time $O(ploylog(MD))$.
\par (2.2) Further appending some qubits in state $| 0 \rangle$, and utilizing $e^{-{\rm i}S_{W}^{1/2} t}$ to perform phase estimation on the second register $|\boldsymbol{\omega}_{j}\rangle$. We can then obtain an approximation to the state $|\Phi_{1}\rangle$ is given by Eq. (\ref{eq:20}).
\par (2.3) Adding another register in the state $|0\rangle$ and applying a controlled rotation is shown in the step (S2.3), then we have the state $|\Phi_{2}\rangle$.
\par (2.4) Undoing phase estimation and removing the register $|\sigma_{k}\rangle$, the rest of the system is in state $|\Phi_{3}\rangle$, which is shown as Eq. (\ref{eq:22}).
\par (2.5)  Finally, performing the projective $|1\rangle \langle 1|$ on the last register of $|\Phi_{3}\rangle$. If measurement succeeds, we trace out this register and obtain the state $|\psi_{T}\rangle$ of the other registers.
\par (3) \emph{Branch and interception}. This step is a key step in our algorithm. Its aims to generate the final desired state $|\psi_{Y}\rangle$ by virtue of the intermediate state $|\psi_{T}\rangle$ in step (2). In other words, to further understand Eq. (\ref{eq:13}) and Eq. (\ref{eq:23}), we need to perform the mapping: ${|\mathbf{v}_{j}}\rangle\mapsto |j\rangle$ on $|\psi_{T}\rangle$ and truncate it to keep the first d terms. Based on this idea, the steps are achieved as follows.
\par (3.1) \emph{phase estimation}. Appending some qubits in the state ${|0\rangle}^{\otimes{L}}$, where $L=log(1/\epsilon_{\lambda})$. And we perform phase estimation of the unitary operation $e^{-{\rm i} \rho t}$ on the second register $|\mathbf{v}_{j}\rangle$ of the intermediate state $|\psi_{T}\rangle$, to obtain the state
     \begin{equation}
        \left|\Psi_{1}\right\rangle=\frac{\sum_{i=1}^{M} \sum_{j=1}^{D} y_{i j} C_{1}|i\rangle\left|\mathbf{v}_{j}\right\rangle\left|\bar{\lambda}_{j}\right\rangle}{\|X\|_{F} \sqrt{p_{1}}},
     \label{eq:26}
     \end{equation}
where $\left|\bar{\lambda}_{j}\right\rangle$ is only used to distinguish from $\left|{\lambda}_{j}\right\rangle$ in step (1), both of which are essentially estimates of the eigenvalue of $e^{-{\rm i} \rho t}$ within the error of $\epsilon_{\lambda}$.
\par (3.2) \emph{Branch}. The purpose of this step is to achieve the shadow principal component and non-principal component branches. At the same time, in order to better obtain the target state $|\psi_{Y}\rangle$, we need to index the first $d$ principal components.
\par Since hosting $d$ indexes require $Q$ qubits and $\left(Q=\lceil log(d) \rceil\right) \leq \left(log(1/\epsilon_{\lambda})=L\right)$, the proof is shown in Appendix \ref{appendix B}. The third register $\left|\bar{\lambda}_{j}\right\rangle$ of the state in Eq. (\ref{eq:23}) is sufficient to satisfy the storage of d indexes. Then, we append another qubit in the state $|0\rangle$ as the last register and perform d unitary operations $U\left( {\lambda_{j}}\right)$.
After it, we can obtain the state
     \begin{equation}
        \begin{split}
            &\left|\Psi_{2}\right\rangle=\\
            &\frac{\sum_{i=1}^{M}C_{1}|i\rangle\left(\sum_{j=1}^{d} y_{i j} \left|\mathbf{v}_{j}\right\rangle|j\rangle|1\rangle+\sum_{j=d+1}^{D} y_{i j} \left|\mathbf{v}_{j}\right\rangle\left|\bar{\lambda}_{j}\right\rangle|0\rangle\right)}{\|X\|_{F}{\sqrt{p_{1}}}}.
        \end{split}
     \label{eq:27}
     \end{equation}
Each $U\left( {\lambda_{j}}\right)$ can be implemented efficiently because the normalized eigenvalues $\lambda_{j}$ are obtained in step (1). At the same time, we assume $\lambda_{j}$ and $j$ have the binary representations of $\lambda_{j}=\lambda_{j}^{L}\lambda_{j}^{L-1}\cdots \lambda_{j}^{2}\lambda_{j}^{1}$ and $j_{Q} j_{Q-1} \cdots j_{2} j_{1}$ respectively. The quantum circuit for $U\left( {\lambda_{j}}\right)$ is shown in Fig. \ref{fig:1}.
\par (3.3) \emph{Interception}. Applying a project measurement $|1\rangle\langle 1|$ on the last register, to see whether it is in the state $|1\rangle$. If the measurement succeed, we discard this register and have the state of other three registers
     \begin{equation}
        \begin{split}
            \left|\Psi_{3}\right\rangle &=\frac{\frac{1}{\|X\|_{F}{\sqrt{p_{1}}}}{\sum_{i=1}^{M}\sum_{j=1}^{d} y_{i j} C_{1} |i\rangle |\mathbf{v}_{j}\rangle |j\rangle}}{\sqrt{\sum_{i=1}^{M}\sum_{j=1}^{d}\left(\frac{1}{\|X\|_{F} \sqrt{p_{1}}} C_{1}\right)^{2}}}\\
            &=\frac{\sum_{i=1}^{M}\sum_{j=1}^{d} y_{i j} |i\rangle |\mathbf{v}_{j}\rangle |j\rangle}{{\sqrt{\sum_{i=1}^{M}\sum_{j=1}^{d}y_{i j}^{2}}}}.
        \end{split}
     \label{eq:28}
     \end{equation}
\par (3.4) \emph{Replacement}. Observing the above states, it is easy to find that we need to discard the register of $|\mathbf{v}_{j}\rangle$ to complete the mapping and obtain the target state $|\psi_{Y}\rangle$.
\par
\begin{figure}
        %\centering
        \includegraphics[width=0.46\textwidth]{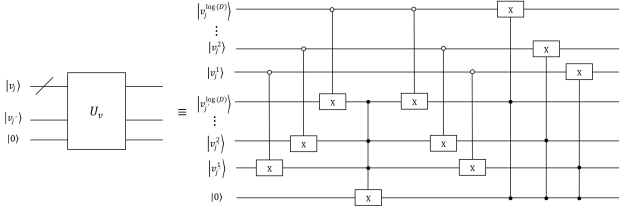}% Here is how to import EPS art
        \caption{Quantum circuit for $U_{\mathbf{v}}$, where ${|\mathbf{v}_{j}\rangle}\equiv|\mathbf{v}_{j}^{log(D)}\mathbf{v}_{j}^{log(D)-1} \cdots \mathbf{v}_{j}^{log(2)} \mathbf{v}_{j}^{1}\rangle$ and $|\mathbf{v}_{j}\rangle \equiv |\mathbf{v}_{j^{\prime}}^{log(D)}\mathbf{v}_{j^{\prime}}^{log(D)-1} \cdots \mathbf{v}_{j^{\prime}}^{log(2)} \mathbf{v}_{j^{\prime}}^{1}\rangle$. Mapping the last register's qubit $|0\rangle$ to state $|1\rangle$ only when $|\mathbf{v}_{j}\rangle= |\mathbf{v}_{j^{\prime}}\rangle$, then implementing the map $|\mathbf{v}_{j}\rangle \mapsto |0\rangle$ by Toffoli gates; Otherwise, the circuit is equivalent to doing nothing.}
    \label{fig:2}
    \end{figure}

 First, selecting a principal component $|\mathbf{v}_{j}\rangle$ from the shadow principal components $|\mathbf{v}_{1}\rangle,|\mathbf{v}_{2}\rangle,\cdots,|\mathbf{v}_{d}\rangle$ which obtained in the step (1). Then, appending a qubit $|0\rangle$ as the last register of the state in Eq. (\ref{eq:28}) and combined with the register of $|\mathbf{v}_{j}\rangle$ to perform the following a unitary operation $U_{\mathbf{v}}$. The quantum circuit for $U_{\mathbf{v}}$ is shown in Figure \ref{fig:2}.
%figure 2
%%%%%equation 29
     \begin{equation}
        U_{\mathbf{v}}:
        \begin{cases}
            |\mathbf{v}_{j}\rangle |\mathbf{v}_{j^{\prime}}\rangle |0\rangle \mapsto |0\rangle |\mathbf{v}_{j^{\prime}}\rangle |1\rangle, &\mathbf{v}_{j}=\mathbf{v}_{j^{\prime}}\\
            |\mathbf{v}_{j}\rangle |\mathbf{v}_{j^{\prime}}\rangle |0\rangle \mapsto |\mathbf{v}_{j}\rangle |\mathbf{v}_{j^{\prime}}\rangle |0\rangle, &\mathbf{v}_{j} \neq \mathbf{v}_{j^{\prime}}
        \end{cases}.
     \label{eq:29}
     \end{equation}
After performed $U_{\mathbf{v}}$, we have the state of whole system,
     \begin{equation}
     \begin{split}
        &\left|\Psi_{4}\right\rangle=\\
        &\frac{\sum_{i=1}^{M} |i\rangle \left( y_{i j} |0\rangle |j\rangle |\mathbf{v}_{j^{\prime}}\rangle |1\rangle + \sum_{j=1,j \neq j^{\prime}}^d y_{i j} |\mathbf{v}_{j}\rangle |j\rangle |\mathbf{v}_{j^{\prime}}\rangle |0\rangle \right)}{\sqrt{\sum_{i}^{M}\sum_{j=1}^{d} y_{i j}^{2}}}.
     \end{split}
     \label{eq:30}
     \end{equation}
Next, discarded the register of $|\mathbf{v}_{j^{\prime}}\rangle$, to get a state
     \begin{equation}
     \begin{split}
        &\left|\Psi_{5}\right\rangle=\\
        &\frac{\sum_{i=1}^{M} |i\rangle \left( y_{i j} |0\rangle |j\rangle  |1\rangle + \sum_{j=1,j \neq j^{\prime}}^d y_{i j} |\mathbf{v}_{j}\rangle |j\rangle  |0\rangle \right)}{\sqrt{\sum_{i}^{M}\sum_{j=1}^{d} y_{i j}^{2}}}.
     \end{split}
     \label{eq:31}
     \end{equation}
Now, we select another principal component form $\{|\mathbf{v}_{j^{\prime}}\}_{j_{\prime}=1}^{d}$, and apply the operation $U_{\mathbf{v}}$ on it and the previously generated state (i.e., the state of $\left|\Psi_{5}\right\rangle$). To repeat this process $d$ times and then obtain the state
     \begin{equation}
        \left|\Psi_{6}\right\rangle=\frac{\sum_{i=1}^{M} \sum_{j=1}^{d} y_{i j} |i\rangle\left|0\right\rangle\left|j\right\rangle |1\rangle}{\sqrt{\sum_{i}^{M}\sum_{j=1}^{d} y_{i j}^{2}}}.
     \label{eq:32}
     \end{equation}
Finally, we discard the second register $|0\rangle$ and the last register $|1\rangle$ to get the state $|\psi_{Y}\rangle.$ It stores the new low-dimensional dataset $\{\mathbf{y}_{i} \in \mathbb{R}^{D} : 1\leq i \leq M\}$ in quantum parallel.
\par
\begin{figure}
        %\centering
        \subfigure[]{
            \includegraphics[width=0.46\textwidth]{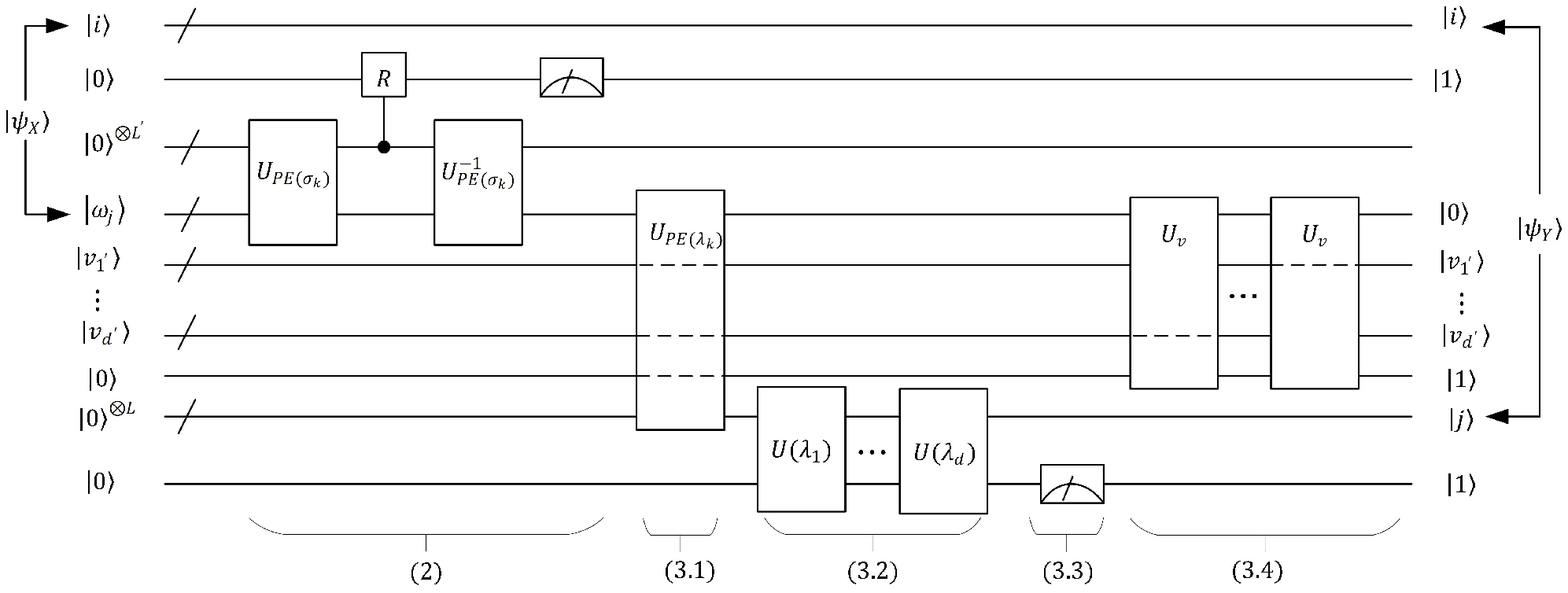}% Here is how to import EPS art
        \label{fig:3a}
        }
           \subfigure[]{
            \includegraphics[width=0.46\textwidth]{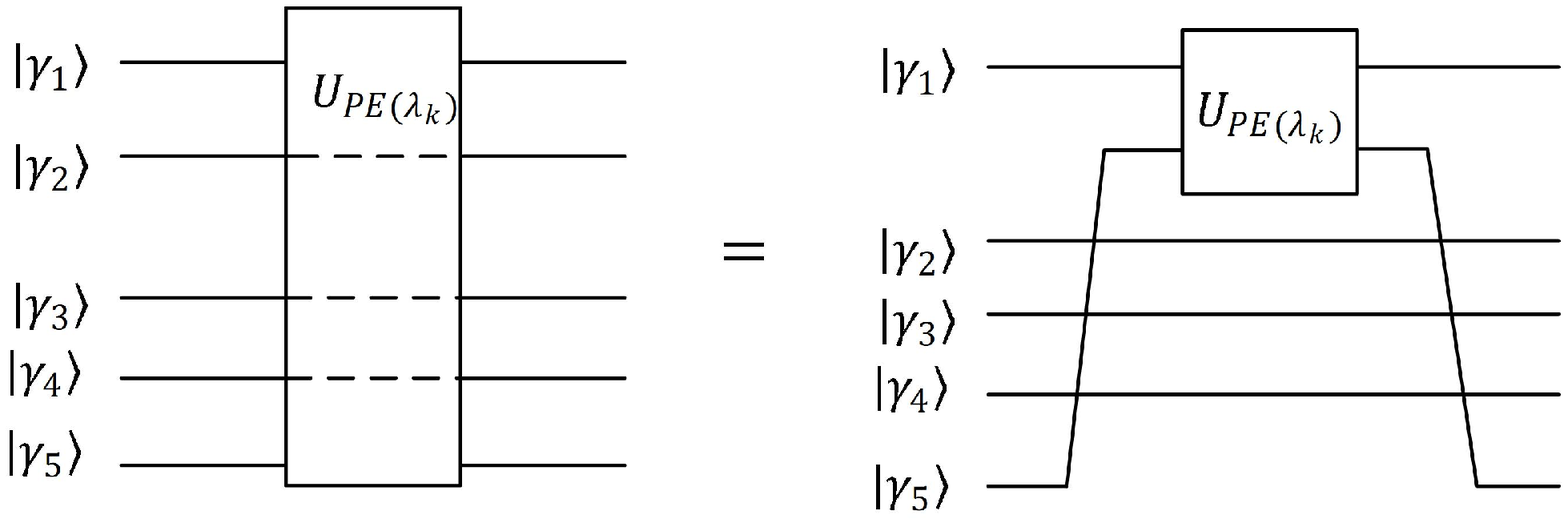}% Here is how to import EPS art
        \label{fig:3b}
        }
        \caption{(a) Quantum circuit for step (2) and step (3). $U_{P E (\sigma_{k})}$, $U_{P E (\lambda_{k})}$ are denoted the quantum circuits phase estimation algorithms for estimating $\sigma_{k}$ and $\lambda_{k}$, respectively. Here, the dotted line indicates that the line does not go through the unitary gate operation, e.g., the unitary operation $U_{P E (\lambda_{k})}$ has not been applied to the dotted line. (b) The actual quantum circuit through the unitary gate $U_{P E (\lambda_{k})}$ in (a).}
        \label{fig:3}
    \end{figure}

 So far we have finished to describe the whole algorithm. Step (1) is the basis, and the other two steps form the main parts of our algorithm. The quantum circuit for step (2) and step (3) are shown in Fig. \ref{fig:3}.
%fig3

%\subsection{\label{sec:3.3}Third-level heading: References and Footnotes}

\section{\label{sec:4}Runtime Analysis}
\par In step (1), according to the conclusion in \cite{ref:SL}, we need $O(1/\epsilon_{\lambda}^{3})$ copies of $\rho$ to ensure that the eigenvalue $\lambda_{j}$ is estimated within the error $\epsilon_{\lambda}$. And, implementing the operator $\rho$ takes time $O( ploylog(M D) k_{\lambda}^{3.5}/\epsilon_{\lambda}^{3} )$. Therefore, we can obtain the first $d$ maximum eigenvalues $\lambda_{1},\lambda_{2},\cdots,\lambda_{d}$ and get one copy of the corresponding eigenvector  $\{|\mathbf{v}_{j}\rangle\}_{j=1}^{d}$ in time $O( d ploylog(M D) k_{\lambda}^{3.5}/\epsilon_{\lambda}^{3})$.
\par Step (2) of the algorithm is to realize the preparation of the intermediate quantum state. Its first step makes us spend time $O(ploylog(M D))$ in constructing the initial state $|\psi_{X}\rangle$ by performing unitary operations $U_{\mathcal{D}}$ and $U_{\mathcal{M}}$. We are easy to find the method in step (2.2) is similar to step (1), which takes time $O(log(M D)k_{\sigma}^{3.5} / \epsilon_{\sigma}^{3})$ to prepare the operator $S_{W}^{1 / 2}$. Furthermore, we need $O(1 / \epsilon_{\sigma}^{3})$ copies of $S_{W}^{1 / 2}$ to construct the operation $e^{-{\rm i} S_{W}^{1 / 2} t}$.  Hence, the time of step (2.2) is $O( ploylog(M D) k_{\sigma}^{3.5}/\epsilon_{\sigma}^{3} )$. Step (2.3) implements the controlled rotation in time $O(log(\epsilon_{\sigma}^{-1}))$. For step (2.4), we consider the probability of success of the post-selection process, the success probability of measurement is
     \begin{equation}
     \begin{split}
        p_{1} &= \frac{\sum_{i=1}^{M}\sum_{j,k=1}^{D} y_{i j}^{2} |\langle \mathbf{u}_{k} | \boldsymbol{\omega}_{j}\rangle|^{2} C_{1}^{2} \sigma_{k}^{2}}{\|X\|_{F}^{2}}\\
        &\geq \frac{\sum_{i=1}^{M}\sum_{j,k=1}^{D} y_{i j}^{2} |\langle \mathbf{u}_{k} | \boldsymbol{\omega}_{j}\rangle|^{2}}{k_{\sigma}^{2}\|X\|_{F}^{2}}\\
        &\geq O(\frac{1}{k_{\sigma}^{2}}),
     \end{split}
     \label{eq:33}
     \end{equation}
because $C_{1}=O(max_{k} \sigma_{k}^{-1})$, $\sum_{k=1}^{D} |\langle \mathbf{u}_{k} | \boldsymbol{\omega}_{j}\rangle|^{2}=1$ and
     \begin{equation}
     \begin{split}
          & \frac{\sum_{i=1}^{M}\sum_{j=1}^{D} y_{i j}^{2} }{\|X\|_{F}^{2}}\\
        = \quad &\frac{\sum_{i=1}^{M}\sum_{j=1}^{D} y_{i j}^{2} }{\sum_{i=1}^{M} \left|\left(\sum_{j=1}^{D} |\boldsymbol{\omega}_{j}\rangle\langle\boldsymbol{\omega}_{j}|\right) \mathbf{x}_{i}\right|^{2}}\\
        \geq \quad &1,
     \end{split}
     \label{eq:34}
     \end{equation}
according to $\sum_{i=1}^{M} \left|\left(\sum_{j=1}^{D} |\boldsymbol{\omega}_{j}\rangle\langle\boldsymbol{\omega}_{j}|\right) \mathbf{x}_{i}\right|^{2} \leq \sum_{i=1}^{M} \sum_{j=1}^{D} \left| \mathbf{x}_{i}^{T} |\boldsymbol{\omega}_{j}\rangle\right|^{2} \left||\boldsymbol{\omega}_{j}\rangle\right|^{2}$. The result of Eq. (\ref{eq:33}) implies that the probability of success is at least $O(k_{\sigma}^{-2})$. In other words, using amplitude amplification \cite{ref:GB}, we are sufficient to get the intermediate state $| \psi_{T} \rangle$ by repeating $O(k_{\sigma})$ times.
\par Step (3) of our algorithm is to obtain the desired state $| \psi_{T} \rangle$, and the time complexity analysis of this step is as follows.
\par Step (3.1) takes time $O(ploylog(MD)k_{\lambda}^{3.5} / \epsilon_{\lambda}^{3})$ to generate the state of Eq. (\ref{eq:26}). In step (3.2), each $CU(\lambda_{j})$ for $j=1,2,\cdots,d$ takes $O(log(1 / \epsilon_{\lambda}))$ elementary gates, so step (3.2) generally takes $O(d log( 1 / \epsilon_{\lambda}))$ time.
\par Further, the probability of successful measurement in step (3.3) is
     \begin{equation}
     \begin{split}
        p_{2} & := \sum_{i=1}^{M}\sum_{j=1}^{d}\left( \frac{ y_{i j}}{\|X\|_{F} \sqrt{p_{1}}} C_{1} \right)^{2}\\
        &= \frac{\sum_{i=1}^{M}\sum_{j=1}^{d} y_{i j}^{2}}{\sum_{i^{\prime}=1}^{M} \sum_{j^{\prime}, k=1}^{D} y_{i^{\prime} j^{\prime}}^{2} \left| \langle \mathbf{u}_{k} | \boldsymbol{\omega}_{j^{\prime}}\rangle \right|^{2} \sigma_{k}^{2}} \\
        &\geq \frac{\sum_{i=1}^{M} \sum_{j=1}^{d} y_{i j}^{2}}{\sum_{i^{'}=1}^{M} \sum_{j^{'}=1}^{D} y_{i^{'} j^{'}}^{2}}.
     \end{split}
     \label{eq:35}
     \end{equation}
According to the fact that $\sum_{k=1}^{D} \left| \langle \mathbf{u}_{k} | \boldsymbol{\omega}_{j^{\prime}}\rangle \right|^{2} = 1$ and $\sum_{k=1}^{D} \sigma_{k}^{2} \leq 1$, let
     \begin{equation}
     \begin{split}
        & \sum_{k=1}^{D}  \left| \langle \mathbf{u}_{k} | \boldsymbol{\omega}_{j^{\prime}}\rangle \right|^{2} \sigma_{k}^{2}\\
         \leq \quad & ( \sum_{k=1}^{D} \left| \langle \mathbf{u}_{k} | \boldsymbol{\omega}_{j^{\prime}}\rangle \right|^{2})( \sum_{k=1}^{D} \sigma_{k}^{2}) \\
        \leq \quad & 1.
     \end{split}
     \label{eq:36}
     \end{equation}
Moreover, it is easy to find
     \begin{equation}
     \begin{split}
        & \frac{\sum_{i=1}^{M} \sum_{j=1}^{d} y_{i j}^{2}}{\sum_{i^{'}=1}^{M} \sum_{j^{'}=1}^{D} y_{i^{'} j^{'}}^{2}}\\
        \geq \quad & 1 - \frac{\sum_{i=1}^{M} |\mathbf{x}_{i}|^{2} \sum_{j=d+1}^{D}\left|| \boldsymbol{\omega}_{j^{\prime}}\rangle \right|^{2}}{\sum_{i=1}^{M} |\mathbf{x}_{i}|^{2} \sum_{j=1}^{D}\left|| \boldsymbol{\omega}_{j^{\prime}}\rangle \right|^{2}} \\
        = \quad & \frac{d}{D}.
     \end{split}
     \label{eq:37}
     \end{equation}
Hence, the lower bound of $p_{2}$ is $O(d/D)$. That is to say, $O(\sqrt{D/d})$ measurements are required to obtain the state of Eq. (\ref{eq:28}) with high probability via amplitude amplification \cite{ref:GB}. And in step (3.4), each $U_{\mathbf{v}}$ needs $O(log D)$ elementary gates, so this step takes $O(d log D)$ time.
\par The time complexity of each step of our algorithm is shown in Table \ref{tab:table1}. To sum up, the total time complexity of the proposed algorithm is
$$\begin{array}{c}
O\left[ploy \log (M D)\left(\left(d+D^{0.5} d^{-0.5}\right) k_{\lambda}^{3.5} \epsilon_{\lambda}^{-3}\right. \right.\\
\left.\left.+D^{0.5} d^{-0.5} k_{\sigma}^{4.5} \epsilon_{\sigma}^{-3}\right)
+D^{0.5} d^{0.5} \log \left(\epsilon_{\lambda}^{-1}\right)\right]
\end{array}$$
This means that if $d = ploylog(D)$, the runtime will be
$$ O[D^{0.5}ploylog(MD)\left(k_{\lambda}^{3.5}\epsilon_{\lambda}^{-3}+ k_{\sigma}^{4.5} \epsilon_{\sigma}^{-3}\right)]$$
Compared with the classical LDA algorithm whose runtime is $O(ploy(M , D))$, the proposed quantum LDA algorithm has exponential acceleration on the number $M$ and shows a quadratic speedup in the original data space dimension $D$.
    \begin{table}
    \caption{\label{tab:table1}The time complexity of each step of our algorithm.}
    \begin{ruledtabular}
    \begin{tabular}{cc}
     Steps& Time complexity\\ \hline
     $(1)$& $O\left( d ploylog(M D) k_{\lambda}^{3.5} / \epsilon_{\lambda}^{3}\right)$ \\
     $(2.1)$& $O\left( \sqrt{(D / d)} \cdot k_{\sigma} ploylog(M D) \right)$ \\
     $(2.2)$& $O\left( \sqrt{(D / d)} \cdot  k_{\sigma} ploylog(M D) k_{\sigma}^{3.5} / \epsilon_{\sigma}^{3} \right)$ \\
     $(2.3)$& $O \left( \sqrt{D / d} \cdot  k_{\sigma} log(1 / \epsilon_{\sigma})\right)$ \\
     $(2.4)$& $O\left( \sqrt{(D / d)} \cdot  k_{\sigma} ploylog(M D) k_{\sigma}^{3.5} / \epsilon_{\sigma}^{3} \right)$ \\
     $(2.5)$& $O\left( k_{\sigma} \right)$ \\
     $(3.1)$& $O\left( \sqrt{(D / d)} \cdot  ploylog(M D) k_{\lambda}^{3.5} / \epsilon_{\lambda}^{3} \right)$ \\
     $(3.2)$& $O \left( \sqrt{D / d} \cdot  d log(1 / \epsilon_{\sigma})\right)$ \\
     $(3.3)$& $O \left( \sqrt{D / d} \right)$ \\
     $(3.4)$& $O \left( d log(D) \right)$ \\
    \end{tabular}
    \end{ruledtabular}
    \end{table}

\section{\label{sec:5}Conclusions}
In this paper, we made a further study on the quantum dimensionality reduction. And we showed a quantum discriminant analysis algorithm for dimensionality reduction. The algorithm can map the data from the high-dimensional space to the low-dimensional space in quantum parallel without changing the original category of the data set. At the same time, we can get a quantum state corresponding to the data set which has been reduced. Compared with the classical LDA algorithm, this quantum algorithm has the ability to achieve exponential acceleration on the number $M$ and quadratic speedup on the dimension $D$ of the original data set. However, how to turn this capability into a stable advantage is worth further exploration. On the other hand, the data after dimensionality reduction exists in the form of a quantum state. So it can be used as the input of other quantum machine learning tasks to overcome the dimension disaster, which has practical significance. In conclusion, this work is conducive to further research on quantum machine learning algorithms in the context of big data.

\begin{acknowledgments}
This work was supported by National Natural Science Foundation of China (Grants No. 61976053 and No. 61772134), Fujian Province Natural Science Foundation (Grant No. 2018J01776), and Program for New Century Excellent Talents in Fujian Province University.
\end{acknowledgments}

\appendix
\section{\label{appendix A}PREPARE THE DENSITY OPERATORS \texorpdfstring{\bm{$S_{W}$}}{$S_{W}$} AND \texorpdfstring{\bm{$S_{B}$}}{$S_{B}} }
\par In this appendix, we give detailed steps for constructing density operators $S_{W}$ and $S_{B}$. As assumed in section \ref{sec:2}, each data vector $\mathbf{x}_{i}$, its class label $c_{i}$ (index $i$ is just a mark for the class of the $i$th data $\mathbf{x}_{i}$) and the centroid $\boldsymbol{\mu}_{c_i}$ are stored as floating-point numbers in QRAM.
\par Furthermore, we consider the $j$th component of $\mathbf{x}_{i}$ can be represented by binary $x_{ij(q-1)} x_{ij(q-2)} \cdots x_{ij(1)} x_{ij(0)}$, where $(q-1),(q-2),\cdots,(0)$ are the marker for the binary bits. Similarly, the $j$th component of centroid $\boldsymbol{\mu}_{c_i}$ has the binary representation $\mu_{c_{i} j (q-1)} \mu_{c_{i} j (q-2)} \cdots \mu_{c_{i} j (1)} \mu_{c_{i} j (0)}$. Then, we may perform the Adder \cite{ref:NM} on the corresponding components of $\mathbf{x}_{i}$ and $-\boldsymbol{\mu}_{c_i}$ to achieve component-wise subtraction. In this way, the difference vector $d_{i j}= x_{i j} - \mu_{c_{i} j}$ is obtained and stored in QRAM. According to the method \cite{ref:SL2,ref:DD}, we consider existing an oracle:
     \begin{equation}
        O_{W}\left( |i\rangle |0\rangle |0\rangle |0\rangle \right) \rightarrow |i\rangle |c_{i}\rangle  | \| \mathbf{x}_{i}-\boldsymbol{\mu}_{c_i}\| \rangle |\mathbf{x}_{i}-\boldsymbol{\mu}_{c_i}\rangle,
     \label{eq:A1}
     \end{equation}
where, $| \| \mathbf{x}_{i}-\boldsymbol{\mu}_{c_i}\| \rangle$ is the value of $\| \mathbf{x}_{i}-\boldsymbol{\mu}_{c_i}\|$ encoded with finite precision on the computational basis and $|\mathbf{x}_{i}-\boldsymbol{\mu}_{c_i}\rangle = \sum_{j=1}^{D} \frac{d_{i j}}{\| \mathbf{x}_{i}-\boldsymbol{\mu}_{c_i}\|} |j\rangle$ is a normalized vector. This oracle could, as an example, be realizable if the data vector components are stored as floating-point numbers in the QRAM, and the sub-norms of the vectors can be estimated efficiently \cite{ref:DD}. Therefore, the oracle $O_{W}$ allows us to construct state
     \begin{equation}
     \begin{split}
          |\varphi\rangle &= O_{W} \left( \frac{1}{\sqrt{M}} \sum_{i=1}^{M}|i\rangle |0\rangle |0\rangle |0\rangle \right)\\
        &= \frac{1}{\sqrt{M}} \sum_{i=1}^{M}|i\rangle |c_{i}\rangle | \| \mathbf{x}_{i}-\boldsymbol{\mu}_{c_i}\| \rangle |\mathbf{x}_{i}-\boldsymbol{\mu}_{c_i}\rangle.
     \end{split}
     \label{eq:A2}
     \end{equation}
By \cite{ref:LG,ref:PK,ref:ANS}, if the norms of the vectors form an efficiently integrable distribution, we will have state
     \begin{equation}
        |\varphi^{\prime}\rangle = \frac{1}{\sqrt{A}} \sum_{i=1}^{M} \| \mathbf{x}_{i}-\boldsymbol{\mu}_{c_i}\| |i> |c_{i}> | \| \mathbf{x}_{i}-\boldsymbol{\mu}_{c_i}\| \rangle |\mathbf{x}_{i}-\boldsymbol{\mu}_{c_i}\rangle,
     \label{eq:A3}
     \end{equation}
where $A=\sum_{i=1}^{M} \| \mathbf{x}_{i}-\boldsymbol{\mu}_{c_i}\|^{2}$.
\par In both cases, we now take the partial trace over the first registers. Then the density matrix of final register $| \mathbf{x}_{i}- \boldsymbol{\mu}_{c_i} \rangle$ can be obtained \cite{ref:IC}:
     \begin{equation}
        S_{W}=\frac{1}{A}\sum_{i=1}^{M} \| \mathbf{x}_{i}-\boldsymbol{\mu}_{c_i}\|^{2} |\mathbf{x}_{i}- \boldsymbol{\mu}_{c_i}\rangle \langle\mathbf{x}_{i}- \boldsymbol{\mu}_{c_i}|.
     \label{eq:A4}
     \end{equation}
The $O_{W}$ efficiently performed makes us construct the Hermitian operator $S_{W}$ in time $O(log(M D))$. And, the within-class scatter matrix $S_W$ can help us run the algorithm efficiently when it is non-singular. Contrary to our expectation, $S_W$ is always irreversible. Therefore, the regularized vector $\boldsymbol{\alpha}$ is usually introduced to make $\boldsymbol{d}_{i}=\mathbf{x}_{i}- \boldsymbol{\mu}_{c_i}+\boldsymbol{\alpha}$ (the component of $\boldsymbol{\alpha}$ is a very small number, which does not affect the classification of the original data and the effect of dimension reduction) \cite{ref:TVB}, that let the density operator $S_{W}$ perfect.
\par Similarly, we assumed $\bar{\boldsymbol{o}}$ (the mean of all data points, $\bar{\boldsymbol{o}}= \frac{1}{M} \sum_{i=1}^{M} \mathbf{x}_{i}$) has been given and stored in the quantum random access memory. Hence, we can cost time $O(log(n D))$ to prepare the between-class scatter density matrix $S_{B}$ in the same way.
     \begin{equation}
        S_{B}=\frac{1}{B}\sum_{c=1}^{n} \| \boldsymbol{\mu}_{c}- \bar{\boldsymbol{o}} \|^{2} |\boldsymbol{\mu}_{c}- \bar{\boldsymbol{o}}\rangle \langle\boldsymbol{\mu}_{c}- \bar{\boldsymbol{o}}|,
     \label{eq:A5}
     \end{equation}
where, $B=\sum_{c=1}^{n} \| \boldsymbol{\mu}_{c}- \bar{\boldsymbol{o}} \|^{2}$. And it is important to note that $S_B$ doesn't have to be invertible in our algorithm.
\par Given the above, we can hence prepare the Hermitian operators $S_{W}$ and $S_{B}$ in time $O(log(M D))$.

\section{\label{appendix B}PROOF OF \texorpdfstring{\bm{$Q \leq L$}}{$Q \leq L$} }
\par In this appendix, we show a proof of $\left(Q=\lceil log(d) \rceil\right) \leq \left(log(1/\epsilon_{\lambda})=L\right)$ in step (3.2), where $d$ denotes the dimension of the have been reduction data set $Y$ and $\epsilon_{\lambda}$ is denoted as the error tolerance of $\lambda_{j}$. The proof is as follows.
\par In step (3.2), the index $j=1,2,\cdots,d$ needs $Q=\lceil log(d) \rceil$ qubits to store and $d$ $(d \neq 0)$ has the binary representation $d = d_{Q} d_{Q-1} \cdots d_{2} d_{1}$. According to the binary division, there have
     \begin{equation}
     \begin{split}
        1 / d &= 1 \div  d_{Q} d_{Q-1} \cdots d_{2} d_{1}\\
        &\Rightarrow
        \begin{cases}
            1 / d = 0_{Q}.0_{Q-1} \cdots 0_{2} 1_{1}, &only \ d_{Q}=1\\
            1 / d \geq 0.0_{Q} 0_{Q-1} \cdots 0_{2} 1_{1}, &others
        \end{cases}.
     \label{eq:B1}
     \end{split}
     \end{equation}
\par Since $\sum_{j=1}^{d} \lambda_{j} \geq \mathcal{S} \approx 1$, each $\lambda_{j}$ scales as $O(1/d)$, and the fact of $\sum_{j=1}^{D} \lambda_{j} = 1$, there exists $\lambda_{j} < 1 / d$ for $j=1,2,\cdots,d$. Furthermore, combined with Eq. (\ref{eq:B1}), it can be known that we need to prepare at least $Q$ qubits to approximate the eigenvalue $\lambda_{j}$ within error $\epsilon_{\lambda}$, that is, $\left(Q=\lceil log(d) \rceil\right) \leq \left(log(1/\epsilon_{\lambda})=L\right)$.

\newpage %Just because of unusual number of tables stacked at end

\begin{thebibliography}{}
%1
\bibitem{ref:CMB}
C. M. Bishop, \textit{Pattern Recognition and Machine Learning (Information Science and Statistics)} (Springer-Verlag, New York, 2006).
%2
\bibitem{ref:GA}
G. Aur\'{e}lien, \textit{Hands-on Machine Learning with Scikit-Learn and TensorFlow} (O'Reilly Media, 2019).
%3
\bibitem{ref:JL}
J. Lever, K.  Martin, and A. Naomi, Nature Methods, \textbf{14}, 641 (2017).
%4
\bibitem{ref:PNB}
P. N. Belhumeur, P. H. Jo\~{a}o, and D. J. Kriegman,  IEEE Transactions on Pattern Analysis and Machine Intelligence, \textbf{19}, 711 (1997).
%5
\bibitem{ref:YQC}
Y. Q. Cheng, K. Liu, J. Y. Yang, Y. M. Zhuang, and N. C. Gu, Proceedings of SPIE - The International Society for Optical Engineering, \textbf{1607}, 85 (1992).
%6
\bibitem{ref:JA}
J. Adcock, E. Allen, M Day, S. Frick, J. Hinchliff, M. Johnson, S. Morley-SHort, S. Palliser, A. Price, and S. Stanisic, arXiv:1512.02900v1 (2015).
%7
\bibitem{ref:PW}
P. Wittek, \textit{Quantum machine learning: what quantum computing means to data mining} (Academic Press, 2014).
%8
\bibitem{ref:MS}
M. Schuld, I Sinayskiy, and F. Petruccione, Physical Review A, \textbf{94}, 022342 (2016).
%9
\bibitem{ref:BJD}
B. J. Duan, J. B. Yuan, Y. Liu, and L. Dan, Physical Review A, \textbf{96}, 032301 (2017).
%10
\bibitem{ref:CHY}
 C. H. Yu, F. Gao, and Q. Y. Wen, An improved quantum algorithm for ridge regression, arXiv:1707.09524v5 (2019).
 %11
\bibitem{ref:JZ}
J. Zhao, Y. H. Zhang, C. P. Shao, Y. C. Wu, G. C. Guo, and G. P. Guo, Physical Review A, \textbf{100}, 012334 (2019).
 %12
\bibitem{ref:AP}
 A. Peruzzo, J. McClean, P. Shadbolt, M. H. Yung, X. Q. Zhou, P. J. Love, A Aspuru-Guzik, and J. L. O'Brien, Nature communications, \textbf{5}, 4213 (2014).
 %13
\bibitem{ref:MC}
M. Cerezo, K. Sharma, A. Arrasmith, and P. J. Coles, arXiv:2004.01372 (2020).
%14
\bibitem{ref:RL}
R. LaRose, A. Tikku, \'{E}. O'Neel-Judy, L. Cincio, and P. J. Coles, npj Quantum Information, \textbf{5}, 1 (2019).
%15
\bibitem{ref:YL}
Y. Liu, D. Y. Wang, S. Xue, A. Huang, X. Fu, X. G. Qiang, P. Xu, H. L. Huang, M. T. Deng, C. Guo, X. J. Yang, and J. J. Wu, Physical Review A, \textbf{101}, 052316 (2020).
%16
\bibitem{ref:SL}
S. Lloyd, M. Mohseni and P. Rebentrost, Nature Physics, \textbf{10}, 108 ( 2014).
%17
\bibitem{ref:CHY2}
C. H. Yu, F. Gao, S. Lin, and J. B. Wang, Quantum Information Processing, \textbf{18}, 249 (2019).
%18
\bibitem{ref:BJD2}
B. J. Duan, J. Yuan, J. Xu, and D. Li, Physical Review A, \textbf{99}, 032311 (2019).
%19
\bibitem{ref:IC}
I. Cong, and L. Duan, New Journal of Physics, \textbf{18}, 073011 (2016).
%20
\bibitem{ref:VG}
 V. Giovannetti, S. Lloyd, and L. Maccone, Physical Review Letters, \textbf{100}, 160501 (2008).
%21
\bibitem{ref:SL2}
 S. Lloyd, M. Mohseni, and P. Rebentrost, arXiv:1307.0411v2 (2013).
%22
\bibitem{ref:IK}
 I. Kerenidis, and A. Prakash, arXiv:1603.08675v3 (2016).
%23
 \bibitem{ref:PR}
 p. Rebentrost, M. Mohseni, and S. Lloyd, Physical Review Letters, \textbf{113}, 130503 (2014).
 %24
\bibitem{ref:MAN}
 M. A. Nielsen and I. L. Chuang, \textit{Quantum Computation and Quantum Information: 10th Anniversary Edition} (Cambridge
University Press, Cambridge, 2010).
%25
\bibitem{ref:JY}
 J. Yang, A. F. Frangi, J. Y. Yang, D. Zhang, and S. Member, IEEE Transactions on Pattern Analysis and Machine Intelligence, \textbf{27}, 230 (2005).
%26
\bibitem{ref:GB}
G. Brassard, P. H${\o}$yer, M. Mosca, and A. Tapp,  arXiv:quant-ph/0005055v1 (2000).
%27
\bibitem{ref:NM}
N. Mikio, and O. Tetsuo, \textit{Quantum Computing: From Linear Algebra to Physical Realizations} (CRC Press, Boca Raton, 2008).
%28
\bibitem{ref:DD}
D. Dervovic, M.  Herbster, P. Mountney, S. Severini, N. Usher, and L. Wossnig, arXiv: arXiv:1802.08227v1 (2018).
%29
\bibitem{ref:LG}
L. Grover, and T. Rudolph, arXiv:quant-ph/0208112v1 (2002).
%30
\bibitem{ref:PK}
P. Kaye, and M. Mosca, arXiv:quant-ph/0407102v1 (2004).
%31
\bibitem{ref:ANS}
A. N. Soklakov, and R. Schack, Physical Review A, \textbf{73}, 012307 (2006).
%32
\bibitem{ref:TVB}
T. V. Bandos, L. Bruzzone, and G. Camps-Valls, IEEE Transactions on Geoence \& Remote Sensing, \textbf{47}, 862 (2009).
\end{thebibliography}

\end{document}